\documentclass[preprint,1p,numbers,sort&compress,merge]{elsarticle}

\usepackage{lineno}
\usepackage{graphicx}  
\usepackage{slashed}   
\usepackage{amsmath}   
\usepackage{subfigure}

\newcounter{JW}

\usepackage[bookmarks=true,colorlinks=true,linkcolor=blue,unicode=true]{hyperref}

\begin{document}

\begin{frontmatter}

\title{Using a nested anomaly detection machine learning algorithm to study the neutral triple gauge couplings at an \texorpdfstring{$e^+e^-$}{e+e-} collider}

\author[1]{Ji-Chong Yang\corref{cor1}}
\ead{yangjichong@lnnu.edu.cn}

\author[1]{Yu-Chen Guo}
\ead{ycguo@lnnu.edu.cn}

\author[1]{Li-Hua Cai}
\ead{984484346@qq.com}

\cortext[cor1]{Corresponding author}

\address[1]{Department of Physics, Liaoning Normal University, Dalian 116029, China}

\begin{abstract}
Anomaly detection algorithms have been proved to be useful in the search of new physics beyond the Standard Model.
However, a prerequisite for using an anomaly detection algorithm is that the signal to be sought is indeed anomalous.
This does not always hold true, for example when interference between new physics and the Standard Model becomes important.
In this case, the search of new physics is no longer an anomaly detection.
To overcome this difficulty, we propose a nested anomaly detection algorithm, which appears to be useful in the study of neutral triple gauge couplings at the CEPC, the ILC and the FCC-ee.
Our approach inherits the advantages of the anomaly detection algorithm been nested, while at the same time, it is no longer an anomaly detection algorithm.
As a complement to anomaly detection algorithms, it can achieve better results on problems that are no longer anomaly detection.
\end{abstract}


\end{frontmatter}


\section{\label{sec:introduction}Introduction}

The search for signals from new physics~(NP) beyond the Standard Model~(SM) is one of the most frontier topics in the field of high energy physics~(HEP)~\cite{johnellis}.
In order to avoid dealing with the huge number of various NP models, a model-independent approach known as the SM effective field theory~(SMEFT) has become popular in the phenomenological studies~\cite{weinberg,*SMEFTReview1,*SMEFTReview2,*SMEFTReview3}.
Since the SM is very successful with only few exceptions~\cite{neutrinomass1,*neutrinomass2,*g2muon,*p5prime,*rdstar1,*rdstar2}, it can be expected that the NP signal that has not yet been observed must be very small.
Moreover, from the SMEFT point of view, the NP signal comes from new interactions, so it can be expected that the kinematic features of NP signals are different.
As a consequence, looking for NP signals is anomaly detection~(AD), which is well suited to machine learning~(ML) algorithms.

ML algorithms have been widely used in HEP~\cite{mlreview,*ml1,*ml2,*ml3,*ml4,*ml5,*ml6,*ml8,*ml10,*ml11,*ml12,*ml13,*ANN1,*ANN2}, for example in the studies that implement the search for NP as AD~\cite{autoencoder1,*autoencoder2,*ml7,*ml9,*guassian,*ad,*ad2,if}.
One of the advantages is that AD is not only model-independent, but also operator-independent.
From the point of view of the operator geometric space, it is not sufficient to study only dimension-6 operators~\cite{convexgeometry}, and phenomenological study of dimension-8 operators has been gaining attention~\cite{d8,bi1,*bi2,*bi3,*aqgc1,*aqgc2,*aqgc3}.
If one has to consider dimension-8 or even dimension-10 operators, the number of operators is very large, e.g., $993$ for dimension-8 and $15456$ for dimension-10~\cite{d8}.
Using an AD algorithm, one can avoid studying the kinematics for specific operators.
Besides, AD algorithms such as the isolation forest~(IF) algorithm~\cite{4781136} are efficient and easy to implement, which are suitable for the rapidly growing volume of data collected at colliders.
The IF algorithm has been shown to be useful in the study of anomalous quartic gauge couplings~(aQGCs) in vector boson scattering processes~\cite{if}.

However, NP signals are not always very different.
For example we find that IF algorithm is not applicable to study the neutral triple gauge couplings~(nTGCs) at the CEPC, the ILC and the FCC-ee.
nTGCs have received a lot of attentions recently because they do not present in the SM and do not receive contributions from dimension-6 operators~\cite{ntgc7,ntgc1,*ntgc4,*ntgc5,*ntgc6,ntgc2,*ntgc3,ntgcloop}.
It has been shown that $e^+e^-$ colliders are suitable for studying nTGCs~\cite{ntgc7,ntgc2,*ntgc3}.
However, when the energy scale is not large, the kinematics of NP signal is not sufficiently distinct from the SM, and the interference becomes important.
In this case, the search for NP signals is no longer AD, and improved algorithms are needed.

In this paper, we propose a nested IF~(NIF) event selection strategy~(ESS), to use the variation of anomaly score to discriminate signal events.
It inherits the advantages of IF, with a transparent mechanism and almost no tunable parameters, it is an unsupervised ML algorithm that does not need to tag the source of events, and it is model-independent and does not depend on the operator to be studied.
We shall emphasize that NIF is no longer an anomaly detection algorithm, therefore it can be expected that NIF can be useful in some problems that cannot be solved by anomaly detection algorithms.
Moreover, not only IF, but in principle any algorithm that gives quantitative measurements of the degree of anomaly for each sample can be nested in our approach.
We find that NIF is useful in the study of nTGCs at the CEPC, the ILC and the FCC-ee.

The remainder of this letter is organized as follows. In Sec.~\ref{sec:problem}, we briefly discuss the problem of IF algorithm in the study of nTGCs, the NIF algorithm with numerical results are presented in Sec.~\ref{sec:mif}, and finally Sec.~\ref{sec:summary} is a summary.

\section{\label{sec:problem}The problem of isolation forest with nTGCs}

The nTGCs receive no tree level contribution neither from the SM nor from the dimension 6 operators~(the contribution at 1-loop level was studied in Ref.~\cite{ntgcloop}).
There are $4$ CP-conserving dimension-8 operators contributing to nTGCs, the Lagrangian is~\cite{ntgc2,*ntgc3,ntgc7}
\begin{equation}
\begin{split}
&\mathcal{L}_{\rm nTGC}=\frac{{\rm sign}(c_{\tilde{B}W})}{\Lambda_{\tilde{B}W}^4}\mathcal{O}_{\tilde{B}W}+\frac{{\rm sign}(c_{B\tilde{W}})}{\Lambda_{B\tilde{W}}^4}\mathcal{O}_{B\tilde{W}}
 +\frac{{\rm sign}(c_{\tilde{W}W})}{\Lambda_{\tilde{W}W}^4}\mathcal{O}_{\tilde{W}W}+\frac{{\rm sign}(c_{\tilde{B}B})}{\Lambda_{\tilde{B}B}^4}\mathcal{O}_{\tilde{B}B},\\
\end{split}
\label{eq.2.1}
\end{equation}
with the operators
\begin{equation}
\begin{split}
 \mathcal{O}_{\tilde{B}W}=i H^{\dagger}\tilde{B}_{\mu \nu} W^{\mu \rho} \left\{D_{\rho },D^{\nu }\right\}H+{\rm h.c.},\;\;
&\mathcal{O}_{B\tilde{W}}=i H^{\dagger} B_{\mu \nu} \tilde{W}^{\mu \rho} \left\{D_{\rho },D^{\nu }\right\}H+{\rm h.c.},\\
 \mathcal{O}_{\tilde{W}W}=i H^{\dagger}\tilde{W}_{\mu \nu} W^{\mu \rho} \left\{D_{\rho },D^{\nu }\right\}H+{\rm h.c.},\;\;
&\mathcal{O}_{\tilde{B}B}=i H^{\dagger}\tilde{B}_{\mu \nu} B^{\mu \rho} \left\{D_{\rho },D^{\nu }\right\}H+{\rm h.c.},\\
\end{split}
\label{eq.2.2}
\end{equation}
where $H$ denotes the SM Higgs doublet, $\tilde{B}_{\mu \nu}\equiv \epsilon_{\mu\nu\alpha\beta} B^{\alpha\beta}$, $\tilde{W}_{\mu \nu}\equiv \epsilon_{\mu\nu\alpha\beta} W^{\alpha\beta}$ with $W_{\mu\nu}\equiv W_{\mu\nu}^a \sigma^{a}/2$ where $\sigma^a$ are Pauli matrices, $c_X$ are dimensionless coefficients.
The $\Lambda _X$ are related with the cutoff scale as $\Lambda _X= \Lambda / |c_X|^{1/4}$.
For simplicity, we define $f_X\equiv {\rm sign}(c_X)/\Lambda ^4_X$.

At tree level, the processes $e^+e^-\to Z\gamma$ can be affected by those operators via $ZV\gamma$ couplings where $V$ is a $Z$ boson or a photon.
It is only necessary to consider the $\mathcal{O}_{\tilde{B}W}$ operator, because $\mathcal{O}_{\tilde{W}W}$ and $\mathcal{O}_{\tilde{B}B}$ do not contribute when $Z$ boson and $\gamma$ are on-shell, and $\mathcal{O}_{B\tilde{W}}$ has the same contribution as the $\mathcal{O}_{\tilde{B}W}$ in this case~\cite{ntgc2,*ntgc3}.
Therefore, in the paper, we only consider the existence of $\mathcal{O}_{\tilde{B}W}$.

\begin{figure}[!htbp]
\centering{
\includegraphics[width=0.9\textwidth]{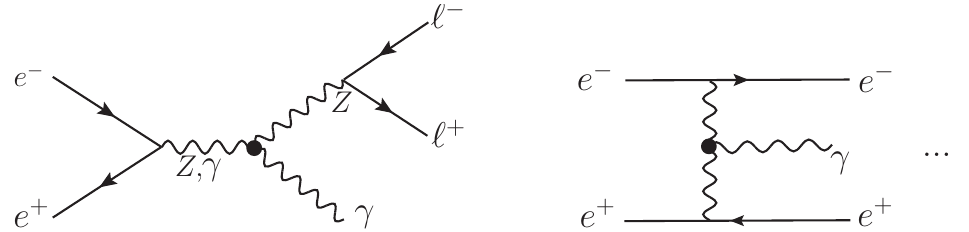}
\caption{\label{fig:feyndiag1}Typical Feynman diagrams for signal events.}}
\end{figure}
\begin{figure}[!htbp]
\centering{
\includegraphics[width=0.9\textwidth]{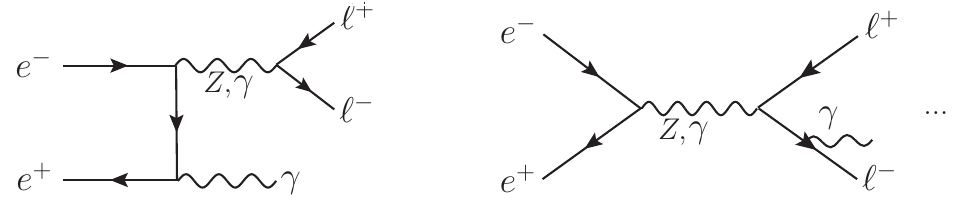}
\caption{\label{fig:feyndiag2}Typical Feynman diagrams for background events.}}
\end{figure}

The dominant contribution from operators in Eq.~(\ref{eq.2.2}) to the process $e^+e^-\to \ell^+\ell^-\gamma$  where $\ell$ is $e$ or $\mu$ is through process $e^+e^-\to Z\gamma$.
Therefore, one can use the process $e^+e^-\to \ell^+\ell^-\gamma$ at $e^+e^-$ colliders to study $\mathcal{O}_{\tilde{B}W}$.
The signal of $\mathcal{O}_{\tilde{B}W}$ is induced by Feynman diagrams shown in Fig.~\ref{fig:feyndiag1}.
In this letter, the events of the processes $e^+e^-\to \mu^+\mu^-\gamma$ and $e^+e^-\to e^+e^-\gamma$ are combined.
The background is the process $e^+e^- \to \ell^+\ell^-\gamma$ in the SM.
The typical Feynman diagrams are shown in Fig.~\ref{fig:feyndiag2}.

The numerical results are studied by using Monte-Carlo~(MC) simulation with the \verb"MadGraph5_aMC@NLO" toolkit~\cite{madgraph,*feynrules}.
A fast detector simulation is applied by using \verb"Delphes"~\cite{delphes} with the CEPC detector card~\cite{cepccard}.
The events are generated at $\sqrt{s}=250\;{\rm GeV}$ which is the energy scale of `Higgs factory' and is close to or covered by the CEPC~\cite{CEPC0,*CEPC1,*CEPC2}, the ILC~\cite{ILC1,*ILC2} and FCC-ee~\cite{FCCee,*FCCee2}.
We use the basic cuts same as the default settings except for $\Delta R_{\ell\ell}$ which is defined as $\sqrt{\Delta \eta_{\ell\ell} ^2+\Delta \phi_{\ell\ell}^2}$ where $\Delta \eta_{\ell\ell}$ and $\Delta \phi_{\ell\ell}$ are the differences between pseudorapidities and azimuth angles of the charged leptons, respectively.
As explained in Ref.~\cite{ntgc7}, the $\Delta R_{\ell\ell}$ is small for an energetic $Z$ boson, so we use $\Delta R_{\ell\ell}>0.2$~\cite{ntgc7,dr02} to avoid losing too much signal events.

\subsection{\label{sec:IF}Isolation forest}

To highlight the signal significance, an ESS based on an IF algorithm has been proposed in Ref.~\cite{if} and was found useful in the study of aQGCs.
The IF algorithm makes use of the fact that the anomalies are `few and different', and can be applied for multi-dimensional data efficiently.
In the IF algorithm, a dimensionless anomaly score~(denoted as $a$) can be calculated for each event where $0<a<1$.
An IF is made of many isolation trees~(ITs).
Consider a dataset in which each piece of data is a multidimensional vector~(denoted as $x^i$), an IT can be constructed using the following procedure.
\begin{enumerate}
  \item Put all points into a node~(denoted as the root node).
  \item Randomly select a node which has not been partitioned yet and contains more than one point~(denoted as $n_0$).
  \item Randomly select a dimension~(denoted as $D$), randomly set a partition value ${\rm min}(x^i_D)< x<{\rm max}(x^i_D)$ where $i$ runs over all points in this node, $x^i_D$ denotes the $D$-th component of $x^i$.
  \item For $n_0$, generate two children nodes~(denoted $n_L$ and $n_R$), move the points in $n_0$ with $x^i_D<x$ into $n_L$, and the others into $n_R$.
  \item Repeat (2) to (4) until every node is either partitioned or contains only one point.
\end{enumerate}
Since random numbers are used to construct an IT, to obtain a stable result, many ITs are constructed to form an IF.
The anomaly score of a point can be obtained as $a=2^{-\bar{L}/c(N)}$, where $\bar{L}$ is the average of the depths of the nodes contain the point over all ITs, $N$ is the number of points in the dataset, $c(N)=2H(N-1)-2(N-1)/N$ is a normalization used to regularize N independently and is the average depth of all nodes in an IT, where $H(N)$ is the harmonic number.
Details on how the anomaly score is calculated can be found in Refs.~\cite{4781136,if}.

To study the signal of nTGCs, $15$ data sets are generated with $\mathcal{L}_{\rm SM}+\mathcal{L}_{\rm nTGCs}$ and with $f_{\tilde{B}W} = -700 + 100 k\;{\rm TeV}^{-4}$ where $0\leq k \leq 14$ are integers.
Each data set consists of $N=500000$ events.
These data sets are in fact those used in Ref.~\cite{ntgc7}, so that we can compare our result with Ref.~\cite{ntgc7}.
We require that there are at least two charged leptons, and the two hardest leptons have the same flavor and different charges.
Besides, we also require that there is at least one photon in the final state.
These requirements are the same as in Ref.~\cite{ntgc7}.
After these requirements there are about $330000$ events left in each data set.
The one piece of data corresponding to each event is a $12$-dimensional vector consisting of components of the 4-momenta of the two hardest charged leptons and the photon.

Note that, both IF algorithm and the NIF algorithm to be introduced are model independent.
Namely, the datasets with $f_{\tilde{B}W}\neq 0$ can be replaced with any dataset without knowing what kind of NP is contained in the dataset under study.
For example, the dataset can be replaced with the dataset obtained in experiment to search for the signal of any possible NP models.
We use the dataset with $f_{\tilde{B}W}\neq 0$ merely to study the performances of IF and NIF on the nTGCs.

\begin{figure}[!htbp]
\centering{
\includegraphics[width=0.6\textwidth]{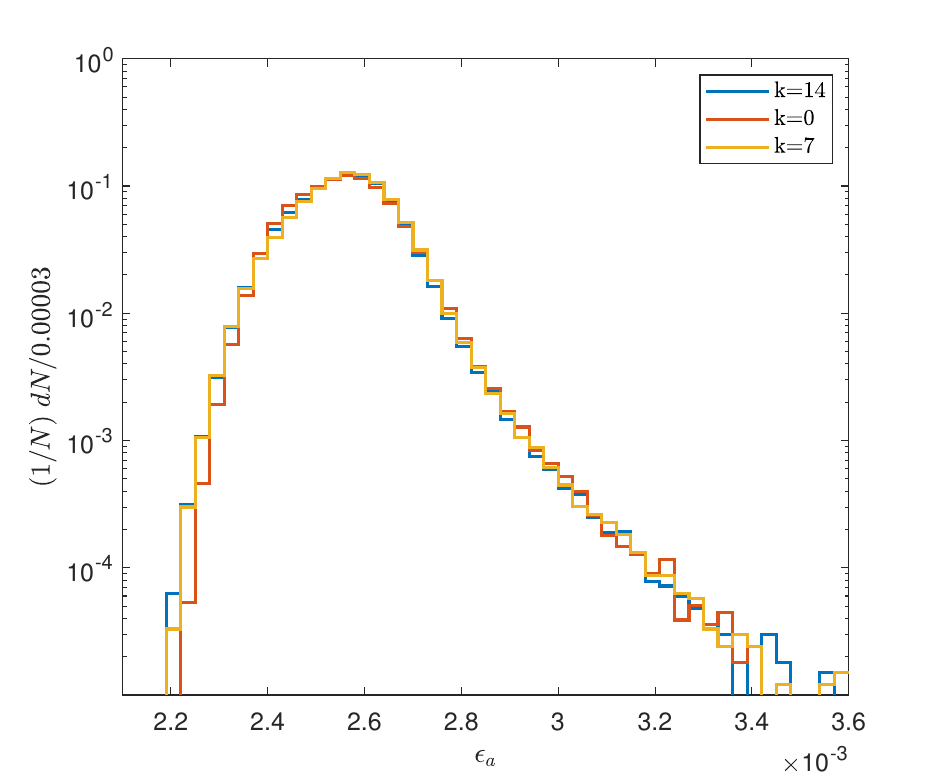}
\caption{\label{fig:error}The normalized distributions of $\epsilon _a$ for $k=0,7,14$ when $n=2000$.}}
\end{figure}
The data preparation is performed by \verb"MLAnalysis"~\cite{Guo:2023nfu}. Once data sets are determined, the only parameter in IF algorithm is the number of isolation trees~(denoted as $n$) which controls the statistical error of the anomaly scores.
In principle, $n$ should be as large as possible as far as the computational power allows.
The standard error of anomaly score~(denoted as $\epsilon _a$) can be calculated for each event.
With $n=2000$, the normalized distributions of $\epsilon _a$ for the events in the data sets with $k=0,7,14$ are shown in Fig.~\ref{fig:error}.
It can be seen that, generally $\epsilon _a<0.004$ and $2000$ trees are sufficient.

\subsection{\label{sec:problemofif}The problem of IF}

\begin{figure}[!htbp]
\centering{
\includegraphics[width=0.6\textwidth]{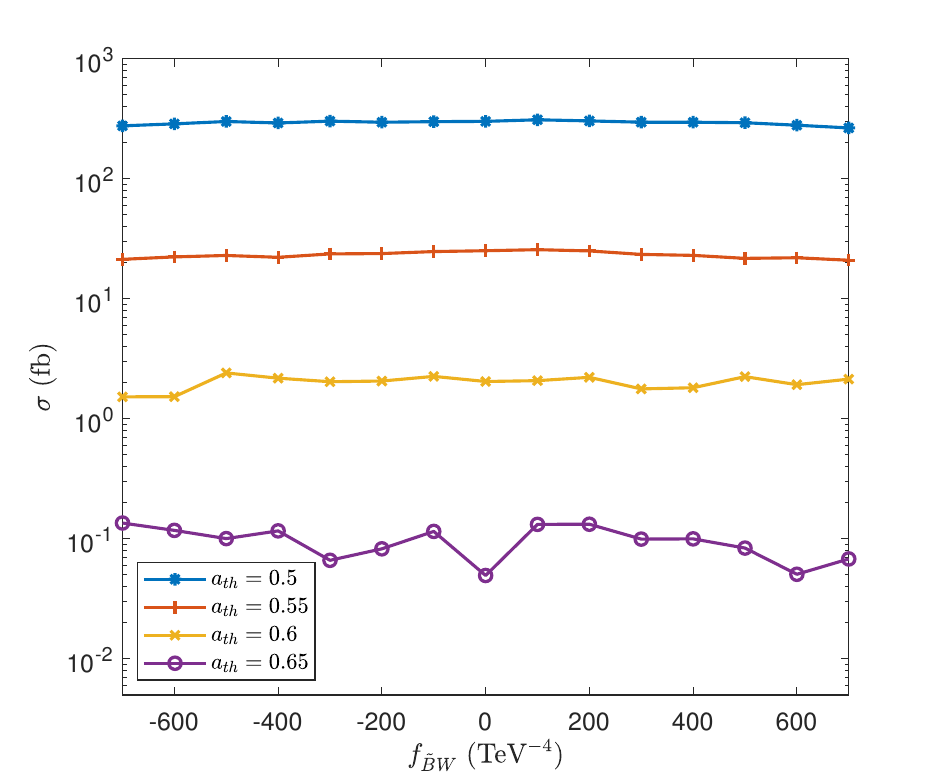}
\caption{\label{fig:csas}The cross-sections as functions of $f_{\tilde{B}W}$ after the ESS that selecting events with $a>a_{\rm th}$.}}
\end{figure}

One can use an ESS to select the event with $a$ greater than a threshold score~(denoted as $a_{\rm th}$) to highlight the signal events.
However, in the case of nTGCs at $\sqrt{s}=250\;{\rm GeV}$, the outcomes are not satisfactory.
The cross-sections with different $a_{th}$ are shown in Fig.~\ref{fig:csas}, no obvious pattern can be seen.
The problem is that when the energy scale is not large, the kinematic features of signal events are not sufficiently distinct from the SM.
In this case the search of the NP signal is no longer a AD problem.

\begin{figure*}[!htbp]
\centering{
\includegraphics[width=0.8\textwidth]{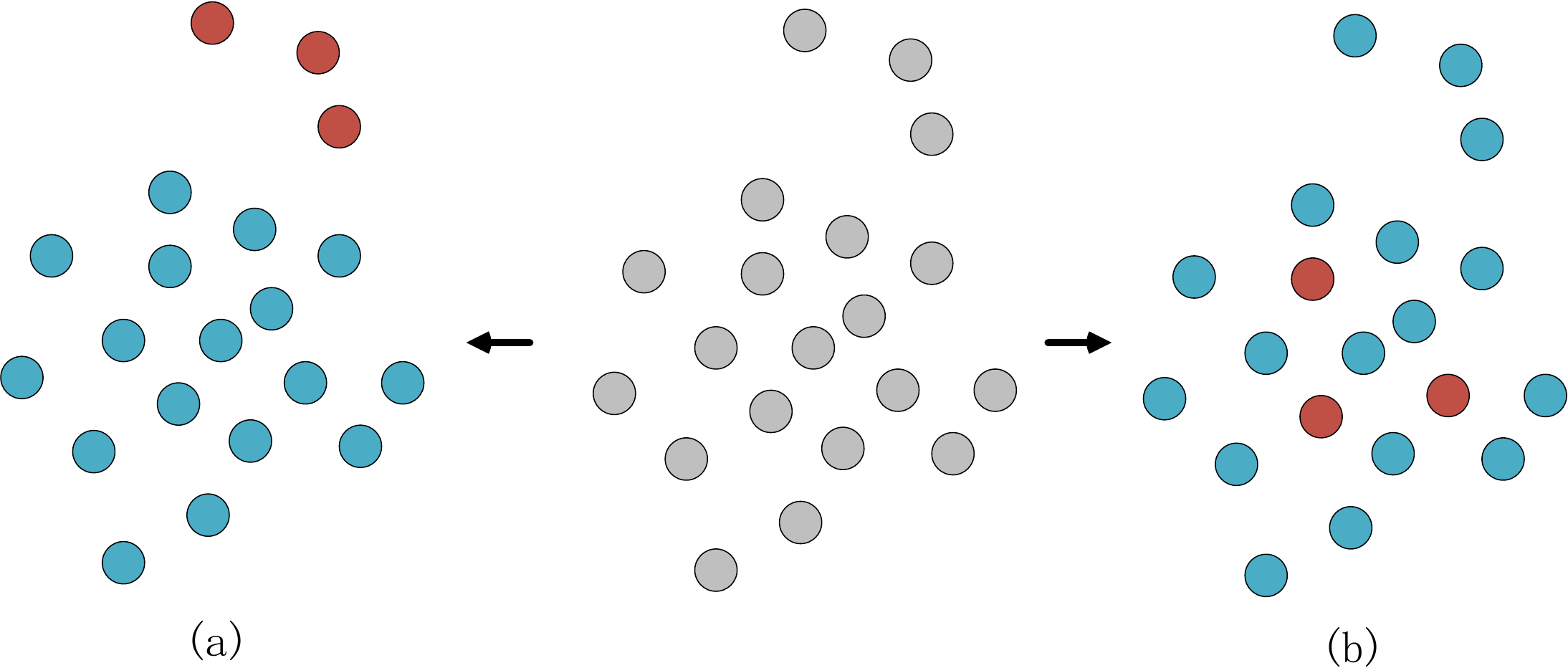}
\caption{\label{fig:problemofad}Two cases of NP signals. The signal points are colored in red and the background points are colored in blue. (a) The signal points are the anomalous points, this is an AD problem. (b) The distribution of signal points overlaps with the background, and this is no longer an AD problem.}}
\end{figure*}

The problem encountered can be depicted in Fig.~\ref{fig:problemofad}.
To illustrate this, it is assumed that the data in the dataset are 2-dimensional vectors and that interference terms are positive or negligible, i.e., the signal events are new points added to the background events.
In addition, assuming that the background events are Gaussian distributed in the phase space, then the rarity of the events can be quantified simply as the distance of the points from the centroid of all events.
In the case of Fig.~\ref{fig:problemofad}~(a), the signal events are anomalous events, and the problem is an AD problem.
In the case of Fig.~\ref{fig:problemofad}~(b), the distribution of the signal events to be searched for overlaps with the background events, and then this is no longer an AD problem.
For the latter, it can be seen that the density of the distribution is increased.
Since anomaly scores are higher in regions with smaller densities, it can be expected that the anomaly scores of some points will decrease when the signal presents.

\section{\label{sec:mif}Nested isolation forest algorithm}

Since the anomaly score calculated by the IF algorithm can be regarded as a representation of the density of events in the phase space, it can be conjectured that anomaly score can also measure the density variation.
In order to achieve this without increasing a large amount of computation, one needs to first build an anomaly score distribution by the SM as a reference.
This can be done by building an IF using the data set from the SM.
Then the NIF ESS can be summarized as follows:
\begin{enumerate}
  \item Using a data set from MC simulation with the SM as a training data set~(denoted as $S_{\rm ref}$), and build an isolation forest, the anomaly score of each event is obtained as $a_{\rm ref}$.
  \item For the data set to be investigated~(from MC simulations or experiments, denoted as $S_{\rm inv}$), build another isolation forest, the anomaly score of each event is obtained as $a_{\rm inv}$.
  \item For each event in $S_{\rm inv}$, find the nearest event in $S_{\rm ref}$, the variation of the anomaly scores for event $i\in S_{\rm inv}$ can be calculated as $\Delta a^i =a^i_{\rm inv}-a^r_{\rm ref}$, where $r\in S_{\rm ref}$ represents the nearest event to $i$.
\end{enumerate}

For this algorithm to work, the distance between the events need to be defined, so that `the nearest event' can be defined.
A systematic way to define the distance known as `earth (or energy) mover’s distance' was introduced in Ref.~\cite{distance}.
Since calculating the distance between events is the most computationally resource-intensive step in this algorithm, we chose a relatively inexpensive way of defining distance.
In this paper, the distance is simply defined as the distance in the phase space as $d=\sqrt{\sum _{ij}(p_j^i-q_j^i)^2}$, where $p_j^i$ and $q_j^i$ are the $i$-th component of the 4-momentum of the particle $j$ in the final state, $j$ runs over the two hardest charged leptons and the photon.
$p$ are 4-momentum of particles in $S_{\rm inv}$ while $q$ are from $S_{\rm ref}$, respectively.

Note that, in this approach, the AD algorithm to be nested is not limited to the IF algorithm.
Any AD algorithm quantifies the degree of anomaly can be nested in this way.
Compared with the AD algorithm to be nested, the additional requirements of this approach are simply the creation of a reference of anomaly scores for $S_{\rm ref}$, and the need to compute distance between events.
Therefore, the nested AD algorithm can inherent the advantages of the algorithms to be nested.


\begin{figure*}[!htbp]
\centering{
\includegraphics[width=0.48\textwidth]{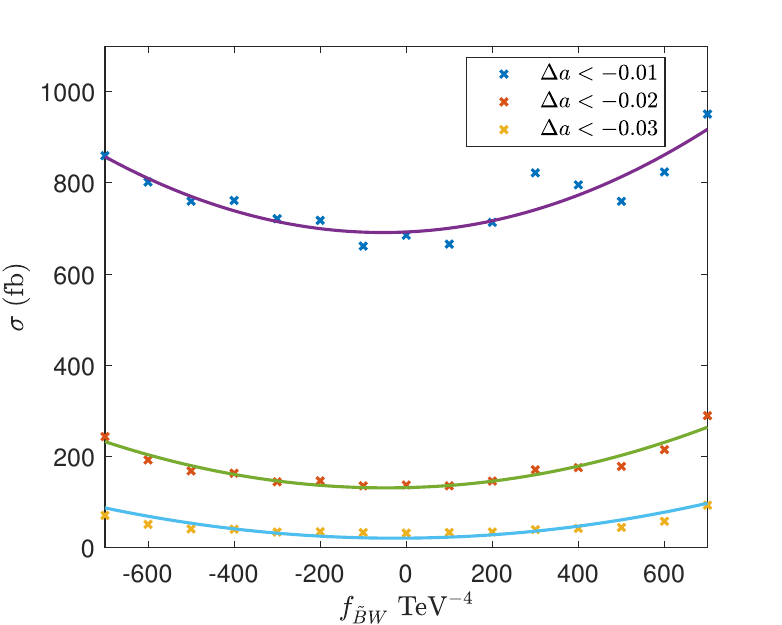}
\includegraphics[width=0.48\textwidth]{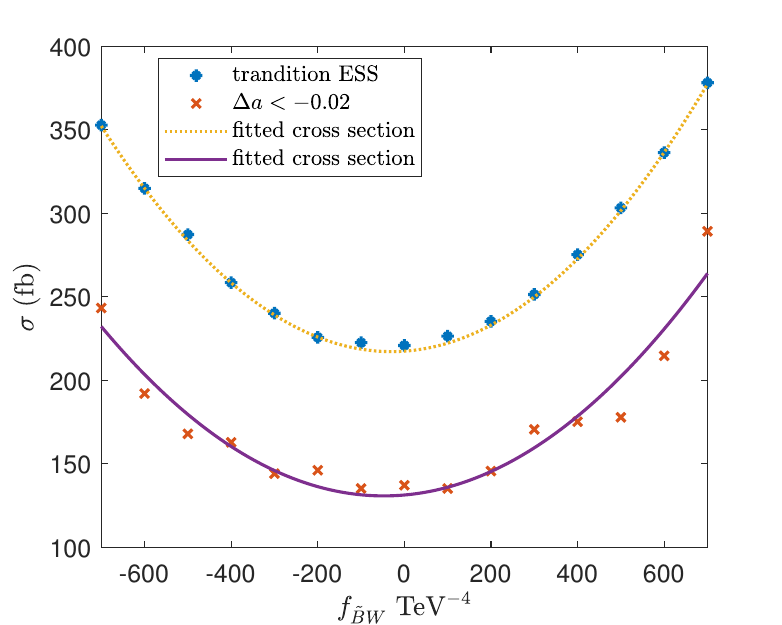}
\caption{\label{fig:cs}The cross-sections as functions of $f_{\tilde{B}W}$ after the traditional ESS and NIF cut.}}
\end{figure*}

Once the variations of the anomaly scores $\Delta a$ are obtained, they can be used to detect the existence of NP models as a complement in the cases such as the nTGCs.

As discussed, when the nTGCs-induced events overlap with those of the SM in the phase space, it mainly leads to a decrease in anomaly score.
Therefore, we select events where $\Delta a$ is less than a certain threshold~(denoted as $\Delta a_{\rm th}$).
This simple strategy is the only one we use in this letter, and is denoted as `NIF cut'.
To build the reference anomaly score distribution, another data set with $N=500000$ are generated with the SM Lagrangian.

The ESS based on kinematic analysis are discussed in Refs.~\cite{ntgc2,*ntgc3,ntgc7}.
We compare the results of NIF cut with the results in Ref.~\cite{ntgc7} which uses an ESS as
\begin{equation}
\begin{split}
 |M_{\ell\ell}-M_Z| < 15\;({\rm GeV})\;\;|\cos \theta _{\gamma}|<0.9,\;\;|\cos \theta _{\ell}|<0.8,\\
\end{split}
\label{eq.3.1}
\end{equation}
where $M_{\ell\ell}$ is the invariant mass of two hardest charged leptons, $\theta _{\gamma}$ is the zenith angle of the photon, $\theta _{\ell}$ is the zenith angle of $\ell^-$ when the $\ell^-$ is boosted into the c.m. frame of two hardest charged leptons whose ${\bf z}$-axis lays along the direction of $\vec{p}_{\ell^+}+\vec{p}_{\ell^-}$.
The cross-sections after traditional ESS~(which is the same as Ref.~\cite{ntgc7}) and after NIF cuts with $\Delta a_{\rm th}=-0.01,-0.02,-0.03$ are shown in Fig.~\ref{fig:cs}.
We find that the cross-section after NIF cut can still be approximated by a bilinear function of $f_{\tilde{B}W}$ within the range of coefficients in use.
Among the results after NIF cut, $\Delta a=-0.02$ is the most consistent with the bilinear function.
Other than that, for the case of $\Delta a=-0.03$, too many signal events are dropped, leading to a smaller signal significance compared with the case of $\Delta a=-0.02$.
For $\Delta a=-0.01$, too many background events are kept.
This can also be seen from the comparison with the traditional ESS that, the cross-section after the NIF cut with $\Delta a=-0.02$ is close to the cross-section after the traditional ESS.

The fitted cross-section after tradional ESS is $\sigma = \left(217.5 +0.018 f_{\tilde{B}W}+ 0.00030 f^2_{\tilde{B}W}\right)\;({\rm fb})$ while after NIF cut is $\sigma = \left(131.4 +0.023 f_{\tilde{B}W}+ 0.00024 f^2_{\tilde{B}W}\right)\;({\rm fb})$.
Compare our result with Ref.~\cite{ntgc7}, it can be seen that, with $\sigma (f_{\tilde{B}W}\neq 0)-\sigma (f_{\tilde{B}W}= 0)$ close to each other~(corresponds to the signal), the $\sigma (f_{\tilde{B}W}= 0)$ using NIF cut is smaller~(corresponds to the background), indicating a compatible or even better signal significance.

Using the signal significance defined as $\mathcal{S}_{stat}=N_{s}/\sqrt{N_s+N_{bg}}$, where $N_{bg}=L\sigma (f_{\tilde{B}W}= 0)$ is the number of events for the SM, $N_{s}=L\left(\sigma (f_{\tilde{B}W}\neq 0)-\sigma (f_{\tilde{B}W}= 0)\right)$ is the number of events beyond the SM with $L$ the luminosity.
For a dataset from the experiment, $N_{bg}=L\sigma _{\rm SM}$ and $N_{s}=L\left(\sigma _{\rm exp}-\sigma _{\rm SM}\right)$ where $\sigma _{\rm exp}$ is the cross-section after NIF cut for the experiment, $\sigma _{\rm SM}$ is the cross-section after NIF cut for the SM by MC simulation.
The expected constraints on $f_{\tilde {B}W}$ at luminosity $L=2\;{\rm ab}^{-1}$ are presented in Table~\ref{Tab:constraints}.
The results of Ref.~\cite{ntgc7} are also listed.
It can be seen that, the NIF cut, which is independent of the operator to be studied, still shows better discriminative ability for $f_{\tilde{B}W}<0$ than the traditional ESS.

\begin{table}
\begin{center}
\begin{tabular}{c|c|c}
 $\mathcal{S}_{stat}$ & traditional ESS & $\Delta a<-0.02$ \\
\hline
 $2$ & $[-85.4, 25.7]$ & $[-114.0, 18.9]$ \\
 $3$ & $[-94.6, 34.9]$ & $[-121.7, 26.6]$ \\
 $5$ & $[-109.8, 50.1]$ & $[-135.1, 40.0]$ \\
\end{tabular}
\end{center}
\caption{\label{Tab:constraints}The constraints on $f _{\tilde{B}W}$ (${\rm TeV}^{-4}$) at $L=2\;{\rm ab}^{-1}$.}
\end{table}

Finally, we would like to emphasize that although the amount of data in the dataset has been irrelevant when calculating the anomaly scores, the larger the amount of data in the SM dataset when building the reference, the more plausible the reference is.
Therefore, it can be expected that the larger the amount of data, the better the NIF will perform.
However, the amount of data is not an tunable parameter because the data is obtained from experiments.
That is, as much data as possible should be collected from the experiments when using the NIF ESS.

\section{\label{sec:summary}Summary}

Although ML algorithms for AD have been widely used in the study of NP, there are cases that the search for NP signals is no longer AD.
This happens when the NP signals are no longer very different, for example when the energy scale is not very large, or when the interference term is important.
In this paper, a nested AD algorithm is proposed for such cases.

The proposed nested anomaly detection algorithm focuses on the variation of the anomaly scores.
Therefore, any anomaly detection algorithm quantifies the degree of anomaly should be able to be nested in our approach.
In this paper, the NIF algorithm is studied.
Such a method inherits the advantages of IF algorithm.
The mechanism behind NIF algorithm is also transparent, there are almost no parameters to be tuned.
The whole procedure is still model independent and operator independent, which works as an automatic ESS.
Moreover, NIF is also an unsupervised machine learning algorithm in the sense that it is not necessary to tag the events.
This is important because it is in principle impossible to specify the origin of an event when there is a negative interference term presented.
Moreover, this unsupervised feature ensures the model-independent, the dataset to be investigated can be a dataset from experiment with the purpose of finding any possible NP model signal that we know or even do not know.

As a complement to the AD algorithms, the problems that a nested AD algorithm targets are those that are no longer AD.
As a case study, the nTGCs at $\sqrt{s}=250\;{\rm GeV}$ are investigated in this paper.
It can be shown that while the IF algorithm is not satisfactory, the NIF is useful in this case.
As an automatic ESS, the NIF can achieve compatible discriminative ability compared to the traditional ESS without knowing at all what operator is under study.
For $f_{\tilde{B}W}<0$, a stronger constraint can be set with the help of an NIF cut.
On the other hand, in the region that the negative interference is dominant, i.e. $\sigma (f_{\tilde{B}W}\neq 0)<\sigma (f_{\tilde{B}W}= 0)$, the traditional ESS is better.
In this region, the pattern how the anomaly score changes is complicated and needs more exploration.
In this case, the NIF cut is still competitive because it achieves a discrimination capacity close to that of traditional ESS without kinematic analysis.

\section*{ACKNOWLEDGMENT}

\noindent
This work was supported in part by the National Natural Science Foundation of China under Grants Nos. 11905093 and 12147214, the Natural Science Foundation of the Liaoning Scientific Committee No.~LJKZ0978 and the Outstanding Research Cultivation Program of Liaoning Normal University (No.21GDL004).

\bibliography{nTGC2}
\bibliographystyle{elsarticle-num}

\end{document}